\documentclass[conference]{IEEEtran}
\IEEEoverridecommandlockouts
\usepackage{cite}
\usepackage{amsmath,amssymb,amsfonts}
\usepackage{algorithmic}
\usepackage{graphicx}
\usepackage{textcomp}
\usepackage{xcolor}
\usepackage{url}
\usepackage{float}
\usepackage[ruled,vlined]{algorithm2e}
\usepackage[T1]{fontenc}
\usepackage{multirow}
\def\BibTeX{{\rm B\kern-.05em{\sc i\kern-.025em b}\kern-.08em
    T\kern-.1667em\lower.7ex\hbox{E}\kern-.125emX}}
    
\begin{document}

\title{Training program on sign language: social inclusion through Virtual Reality in ISENSE project}

\author{\IEEEauthorblockN{Alessia Bisio}
\IEEEauthorblockA{\textit{Dept. of Mechanical} \\
\textit{and Aerospace Engineering} \\
\textit{Politecnico di Torino}\\
Torino, Italy \\
s289321@studenti.polito.it}
\and
\IEEEauthorblockN{Enrique Yeguas-Bolívar }
\IEEEauthorblockA{\textit{Dept. of Computing} \\ 
\textit{and Numerical Analysis} \\
\textit{University of Cordoba}\\
Córdoba, Spain \\
eyeguas@uco.es}
\and
\IEEEauthorblockN{Pilar Aparicio-Martínez}
\IEEEauthorblockA{\textit{Dept. of Nursing, Physiotherapy} \\
\textit{and Pharmacology} \\
\textit{University of Cordoba}\\
Córdoba, Spain \\
n32apmap@uco.es}
\and
\IEEEauthorblockN{María Dolores Redel-Macías}
\IEEEauthorblockA{\textit{Dept. of Rural Engineering} \\
\textit{University of Cordoba}\\
Córdoba, Spain \\
mdredel@uco.es}
\and
\IEEEauthorblockN{Sara Pinzi}
\IEEEauthorblockA{\textit{Dept. Physical Chemistry and Applied Thermodynamics} \\
\textit{University of Cordoba}\\
Córdoba, Spain \\
qf1pinps@uco.es}
\and
\IEEEauthorblockN{Stefano Rossi}
\IEEEauthorblockA{\textit{DEIM, Industrial Engineering} \\
\textit{Università della Tuscia}\\
Viterbo, Italy \\
stefano.rossi@unitus.it} 
\and
\IEEEauthorblockN{Juri Taborri}
\IEEEauthorblockA{\textit{DEIM, Industrial Engineering} \\
\textit{Università della Tuscia}\\
Viterbo, Italy \\
juri.taborri@unitus.it} } 

\maketitle

\begin{abstract}

Structured hand gestures that incorporate visual motions and signs are used in sign language. Sign language is a valuable means of daily communication for individuals who are deaf or have speech impairments, but it is still rare among hearing people, and fewer are capable of understand it.
Within the academic context, parents and teachers play a crucial role in supporting deaf students from childhood by facilitating their learning of sign language.
In the last years, among all the teaching tools useful for learning sign language, the use of Virtual Reality (VR) has increased, as it has been demonstrated to improve retention, memory and attention during the learning process. The ISENSE project has been created to assist students with deafness during their academic life by proposing different technological tools for teaching sign language to the hearing community in the academic context. As part of the ISENSE project, this work aims to develop an application for Spanish and Italian sign language recognition that exploits the VR environment to quickly and easily create a comprehensive database of signs and an Artificial Intelligence (AI)-based software to accurately classify and recognize static and dynamic signs: from letters to sentences.

\end{abstract}


\section{Introduction}

Sign language utilizes structured hand gestures, visual motions, and signs to reinforce communication and convey information in everyday conversations. Sign language is essential for daily interaction within the deaf and speech-impaired community. However, sign language is rare among hearing people, and fewer are capable of understand it. This poses a communication barrier between the deaf community and the rest of society, a problem yet to be fully solved \cite{1}. 

Accurate detection and recognition systems can bridge the gap between sign language users and those who do not understand sign language, facilitating smoother interactions, improving the inclusivity, indipendence and autonomy of deaf individuals.
By providing real-time interpretation of sign language, technology can empower individuals to access information, engage in educational and professional settings, and interact with the digital world more independently, reducing their reliance on interpreters or intermediaries.
These systems can provide real-time feedback, support interactive lessons, and enable teachers and students to communicate effectively, enhancing the overall learning experience.
In fact, regarding deaf students, parents and teachers play an important role by assisting them since childhood in learning sign language to communicate with others; for this reason it is very useful to have interactive learning tools and assistive technologies to help them in this task~\cite{2}.

One of the technologies used in the last years in this context is Virtual Reality (VR), that has several benefits including enhanced empathy, improved communication skills, increased understanding of diversity, and the creation of inclusive learning environments. VR provides immersive experiences that enable students to step into others' perspectives, fostering empathy, promoting a sense of belonging and encouraging social interaction and cooperation among students~\cite{3}.


This paper presents a standalone VR and AI-based recognition system for the Spanish and Italian sign language. 
The main novelty of our system is its aim to serve as a starting point for scalable gesture recognition in sign languages, combining both static signs (letters) and dynamic signs (words or phrases), as required by the deaf community.
The software has been developed under the training program on national sign language for social inclusion through VR in ISENSE, an European project that aims at implementing supporting services to assist students with deafness during their academic life, with a particular focus on the enrolment phase, to increase the accessibility of deaf students in the academic context.

In the upcoming sections, we will delve into the current state of the art, followed by an elucidation of 
the framework employed. Subsequently, we will outline the primary objectives, methods, and experimental design. 
Finally, we will discuss the results and present the conclusion and future work.

\section{Related works}

In recent years, there has been a growing interest in harnessing the power of virtual reality (VR) technology to promote social inclusion and enhance communication accessibility for individuals with diverse needs. On one hand, researchers and developers have explored the potential of VR applications to create immersive environments that foster inclusivity, enabling people of all abilities to participate in shared experiences. On the other hand, significant progress has been made in the development of sign language recognition systems, leveraging computer vision and machine learning techniques to bridge the communication gap between deaf individuals who use sign language and those who do not. 

\subsection{Virtual Reality for Social Inclusion}
Virtual reality systems have emerged as promising tools for promoting social inclusion among various populations. With their immersive and interactive capabilities, VR systems offer unique opportunities to create inclusive environments and bridge social barriers. The VR for Good program~\cite{VRForGood} by Meta is an initiative aimed at utilizing virtual reality technology for positive social impact. The program offers resources, mentorship, and funding to participants, enabling them to leverage VR as a powerful storytelling medium. Projects span various topics, allowing users to experience different perspectives and drive empathy. 

In recent years, there has been a growing interest in leveraging VR technology to foster social inclusion, particularly among diverse groups such as neurodiverse individuals, marginalized communities and individuals with disabilities, that include hearing impaired people too.  
The study presented in~\cite{4} has as main purpose to create virtual reality (VR) games that cater to the diverse sensory needs and preferences of neurodiverse individuals, considering factors such as audio, visual, and haptic stimuli. The goal is to enhance the gaming experiences and inclusivity for neurodiverse children, allowing them to engage and interact with VR games in a way that accommodates their unique sensory profiles. 
In various social contexts individuals who are deaf or hard of hearing often face barriers that hinder their full participation and enjoyment; for example, in~\cite{5} they discuss about the potential of VR as a tool to provide real-time captioning and sign language interpretation during theater shows, thereby improving the theater experience for the target audience. 

\subsection{Sign Language Detection}
In the past years, most works related to sign language recognition did not exploit Virtual Reality (VR), but only simple acquisition systems and Artificial Intelligence (AI) based algorithms for sign classification. An example of automatic sign language recognition is presented in~\cite{6}. With a dataset of 20 different Italian gestures captured using a Microsoft Kinect, the model comprises two Convolutional Neural Networks (CNNs) for hand and upper body feature extraction, followed by concatenation and classification using an Artificial Neural Network (ANN) with LCN and ReLU activation.
In~\cite{7} the authors employed the Kinect sensor to capture depth data of 25 signs of German Sign Language, then they implemented a two-step classification approach: first, they used a depth-based feature extraction method to extract relevant features, then a k-nearest neighbors (k-NN) for gesture classification, evaluating the performance of their approach using cross-validation techniques.

In the last years, instead, the use of VR for teaching purposes has increased, also for learning sign language. Recent studies confirmed that VR improves retention, memory, attention and motivation during learning, because the immersive nature of VR allows the interaction with the virtual world and the creation of a creative and engaging learning environment~\cite{8}. Most of the works regarding sign language learning using VR exploit external acquisition systems to acquire videos of the signs and AI algorithms to perform the recognition in a VR environment. 
The recognized signs are usually the letters of the alphabet, which still falls short of the real needs of the deaf community.
In~\cite{8} they combine the avatar’s hand and body movements recorded using motion capture devices with a synchronized video of the face captured by a head-mounted camera. The avatar’s movements are then reconstructed in the VR environment and used to teach sign language, but live sign recognition is not performed.
A study on the recognition of American Sign Language (ASL) gestures in a VR environment using the Leap Motion controller with two IR cameras to track hand gestures in real-time is presented in~\cite{9}. 
A combination of Hidden Markov Models (HMMs) and Support Vector Machines (SVMs) is used to classify signs. In this case, the different gestures correspond to the letters of the language.
Unlike the works presented so far, our VR-based recognition system is scalable in terms of the number of gestures required to define a specific sign, allowing for the recognition of both static signs such as letters and dynamic signs such as words or phrases.

\section{Framework and Objectives}

\begin{figure}[h]
  \centering
  \includegraphics[scale=0.66]{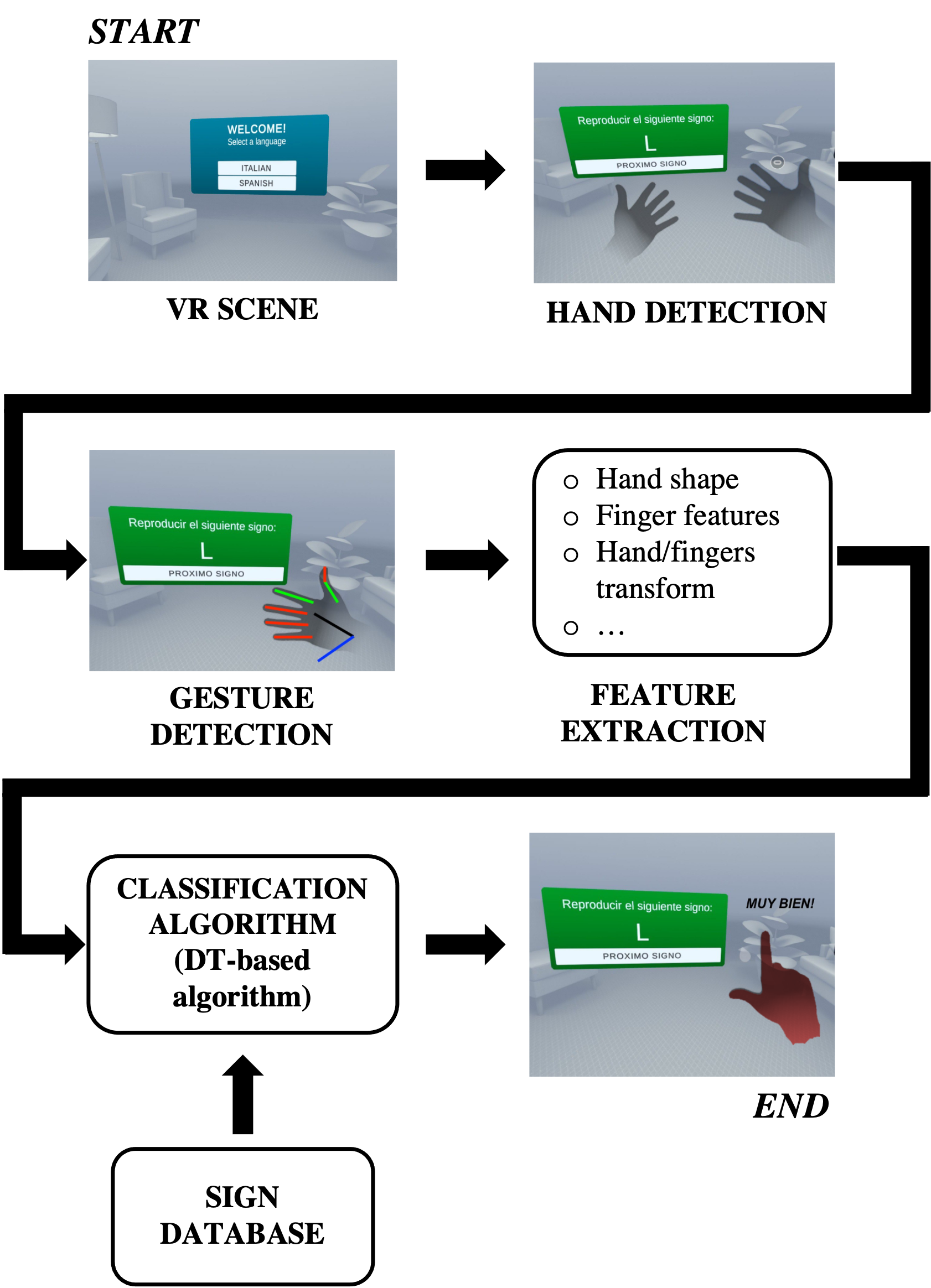}
  \caption{Overview of the sign detection system}
  \label{fig:1}
\end{figure}

One of the objectives of the ISENSE project is to develop a tool that could facilitate and increase sign language learning in the academic context for the hearing community (students, professors, technicians, relatives and friends of deaf students), exploiting the advantages and benefits of VR applications. To achieve this, we developed a VR application for Oculus Meta Quest 2~\cite{10} for the recognition of Spanish and Italian sign language. We exploited the hand tracking done by the four cameras integrated into the Oculus Meta Quest 2 to carry out an on-the-fly sign recognition.
In the application, there are different VR scenarios by which users can add new signs to the database, learn new signs and test the acquired skills.
\begin{enumerate}
    \item \textbf{Creation of the database}. The user can add new signs to the database. This function is intended to be used by expert sign language speakers. First static poses are saved (e.g., alphabet and digits), then dynamic signs of words and simple sentences (e.g., colors, emotions, habits and terms related to academic context) are intro- duced in the database. To perform the dynamic detection, two or three static poses for each sign are saved and the translation between them is then computed. For each pose the arrangement of the fingers in the hand’s position is saved.
    \item \textbf{Learning new signs}. The gesture saved in the first modality are reloaded and rendered into the avatar’s hands. Users can reproduce and learn the sign by putting their hands to fit inside the virtual hands of the avatar. Then the system calculates the accuracy of the handshape and validates in real-time, using Machine Learning techniques, the matching quality between the avatar and the user hands.
    \item \textbf{Test the acquired skills}. Finally, users can test the learned skills, by trying to sign a word from the dictionary without the help of the avatar. The application will give feedback according to the quality of the reproduced sign.
As presented in the overview of the sign detection system (Fig.~\ref{fig:1}), the matching algorithm used for the classification of signs is a decision tree-based algorithm that takes as inputs the features extracted from each sign (e.g., hand shape position, fingers position, the distance between fingers...) in real-time and the signs already saved in the database. Therefore, the AI module performs the matching between the two different inputs and classifies the gesture made by the user.

\end{enumerate}

\section{Methods and Experimental Design}

The methods employed in this study aim to accurately detect and interpret hand gestures in real-time. In this section, we provide a comprehensive overview of the methods used, including data collection, preprocessing, feature extraction, and the utilization of a decision tree algorithm for gesture classification.

\subsection{Machine Learning algorithms for Sign Language classification}
There exist various Machine Learning (ML) algorithms suitable for hand gesture recognition in sign language, with several notable options for gesture classification including:
\begin{enumerate}
    \item \textbf{Convolutional Neural Networks (CNN)}: well-suited for gesture recognition thanks to their hierarchical representation (from simple to more complex patterns in subsequent layers); Support Vector Machine (SVM), a supervised learning algorithm that has good generalization capability (can handle complex decision boundaries) is effective in cases where the data is not linearly separable and is robust to noise and outliers~\cite{11,13}.
    \item \textbf{k-Nearest Neighbour (kNN)}: a simple yet effective algorithm that works well with small training sets and does not make strong assumptions about the underlying data distribution. KNN is easy to understand and implement, making it suitable for various classification tasks~\cite{11}.
    \item \textbf{Decision Tree (DT)-based algorithms}: it can be a simple Decision Tree or a Random Forest, an ensemble learning method that combines multiple decision trees~\cite{11,12}.
    The Machine Learning algorithm for hand gesture recognition implemented in this work is a Decision Tree Classifier (DTC). This model is particularly useful for this application because it offers an intuitive and easy-to-understand representation of the decision-making process; it has fast training and prediction times compared to many other classification algorithms, thanks to the efficient splitting of feature values based on specific criteria; DTCs can also handle missing values, outliers and noisy data, making decisions based on majority voting; DTCs inherently perform feature selection by evaluating the importance of features in the classification process: important features are placed closer to the root of the tree, while less significant features are placed deeper in the tree. This feature selection capability helps in identifying relevant attributes for classification. 
    Other advantages of a DT-based model are the ability to capture nonlinear relationships, versatility in handling different data types, and scalability to large datasets~\cite{14}.
    In the construction of a DTC there are different factors and criteria taken into account. In our work, in particular, we set the stopping criteria to determine when to stop growing the tree and the splitting criterion for selecting the best attribute to split the data at each node. The algorithm uses information gain or entropy as the criterion for selecting the best attribute: the one with the lowest entropy and the highest information gain is selected as it provides the most significant reduction in entropy and maximizes the information gained about the class labels.
    
\end{enumerate}

\begin{algorithm}
\SetAlgoLined
\DontPrintSemicolon

posesList = CaptureSignPose(typeOfGesture)\;

\ForEach{pose in posesList}{
   \ForEach{bone in fingerBonesList}{
       handsDataToSave.Add(bone.position)\;
   }
}

SaveDataToJson(handsDataToSave, posesList, filePath)\;

listOfSavedSigns = LoadHandGestures(filePath, posesList)\;

savedHandsFeatures = CalculateFeatures(listOfSavedSigns, posesList)\;

decisionTree = CreateDecisionTree(savedHandsFeatures)\;

currentHandsData = new HandsData(posesList)\;

\ForEach{pose in posesList}{
   \ForEach{bone in fingerBonesList}{
       currentHandsData.Add(bone.position)\;
   }
}

currentHandsFeatures = CalculateFeatures(currentHandsData, posesList)\;

prediction = new Dictionary<Gesture, Confidence>()

\ForEach{savedGesture in listOfSavedSigns}{
   prediction[savedGesture] = decisionTree.Classify(currentHandsFeatures, savedGesture)\;
}

PerformAction(prediction)\;

\caption{Real-Time Hand Gesture Recognition for Sign Language (Multiple Positions)}
\label{alg:gesture-recognition}
\end{algorithm}

The pseudocode presented in Algorithm~\ref{alg:gesture-recognition} aims to perform real-time hand gesture recognition for sign language using multiple hand positions. The algorithm consists of twelve different steps:

1. Obtain the number of poses involved in the sign gesture by capturing the sign pose for the specified type of gesture. Signs can be static, represented with a single pose, or dynamic, represented with the translation between two or more static poses. 

2. Iterate over each pose in the poses list and collect the position data of finger bones for each hand.

3. Save the hand gestures captured into a defined structure.

4. Load the previously saved hand gestures from the database, considering the specific poses involved in the sign gesture. 

5. Calculate the hand features based on the collected hand positions for each pose involved in the sign gestures previously loaded. 

6. Create a decision tree classifier using the loaded hand gestures as training data. 

7. Initialize the data structure to store the current hand positions recorded in real-time for each pose involved in that specific sign gesture.

8. Iterate over each pose in the poses list and collect the position data of finger bones for both hands and save the current hand position.

9. Calculate the hand features based on the collected hand positions for each pose of the current sign gesture reproduced. 

10. Initialize a dictionary to store the prediction results, associating each gesture with its confidence level.

11. Perform real-time classification for each saved gesture by utilizing the decision tree classifier and the calculated hand features.

12. Based on the predicted gesture, take appropriate actions or responses.



\subsection{Dataset}
For this study, a custom dataset was created, encompassing a total of 50 sign variations per language: alphabet letters, words, and sentences. The dataset specifically focuses on capturing static signs representing each letter of the alphabet, as well as two key pose signs for general concepts such as colors, measurements, emotions, characteristics, and more. Additionally, it incorporates three crucial pose signs for phrases, encompassing commonly used idiomatic expressions, greetings, questions, and popular phrases.

To ensure the dataset's accuracy and diversity, extensive efforts were made to compile a variety of sign examples. These examples were sourced from an array of videos, extensively referenced from the vast collection available at~\cite{15}. This methodology ensured that the dataset encompassed a wide range of sign variations, including variations in hand shape, movement and orientation.

To enhance the dataset's richness, each sign was recorded by multiple users, resulting in a total of 10 instances for each sign. This approach took into account the natural variations that may occur due to individual signing styles and hand sizes, thereby capturing a more comprehensive representation of the sign language system.

For each pose a set of 22 features were calculated, including the euclidean distances among all the key joints of the hand skeleton's bones (distance tip-base of a finger, distance between tips of different fingers and distance tip-intermediate or tip-proximal). 

The dataset was divided into two sets: a training set and a test set with a 70\%-30\% distribution.

\begin{figure}
  \centering
    \includegraphics[width=0.60\linewidth]{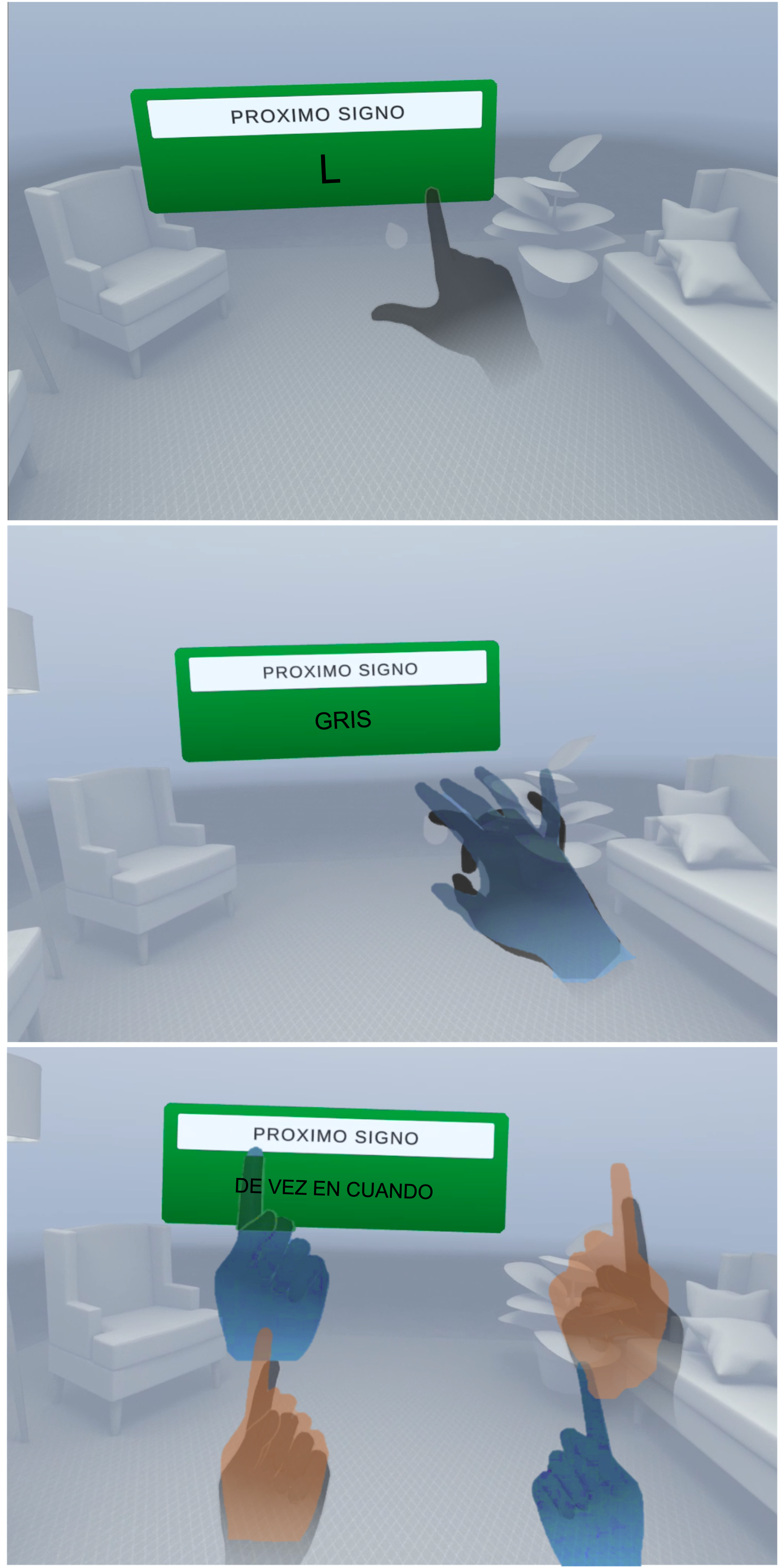}
  \caption{Sign language gestures categorized: alphabet signs, words and sentences.}
  \label{fig:signs}
\end{figure}

The ultimate goal of the experimentation is to achieve the optimal decision tree model and configuration for real-time prediction of a given sign (whether it is an alphabet letter, word, or sentence) in a virtual reality (VR) setting (see Fig.~\ref{fig:signs}). Prior to the classification process, a careful selection of the best parameters has been conducted, including the criterion, maximum depth, minimum samples split, and minimum samples leaf.

To assess the performance of the final decision tree model, it will be compared across different outputs based on accuracy and F1-Score rate metrics. To ensure a fair comparison, the experimentation results for each output and configuration will be obtained using a 10-fold cross-validation approach. This technique involves dividing the dataset into ten subsets, training and evaluating the model ten times using different combinations of training and testing data, and then averaging the results to provide a comprehensive evaluation of the model's effectiveness.

    
\section{Results and Discussion}

\begin{figure*}
  \centering
  \includegraphics[width=0.7\textwidth]{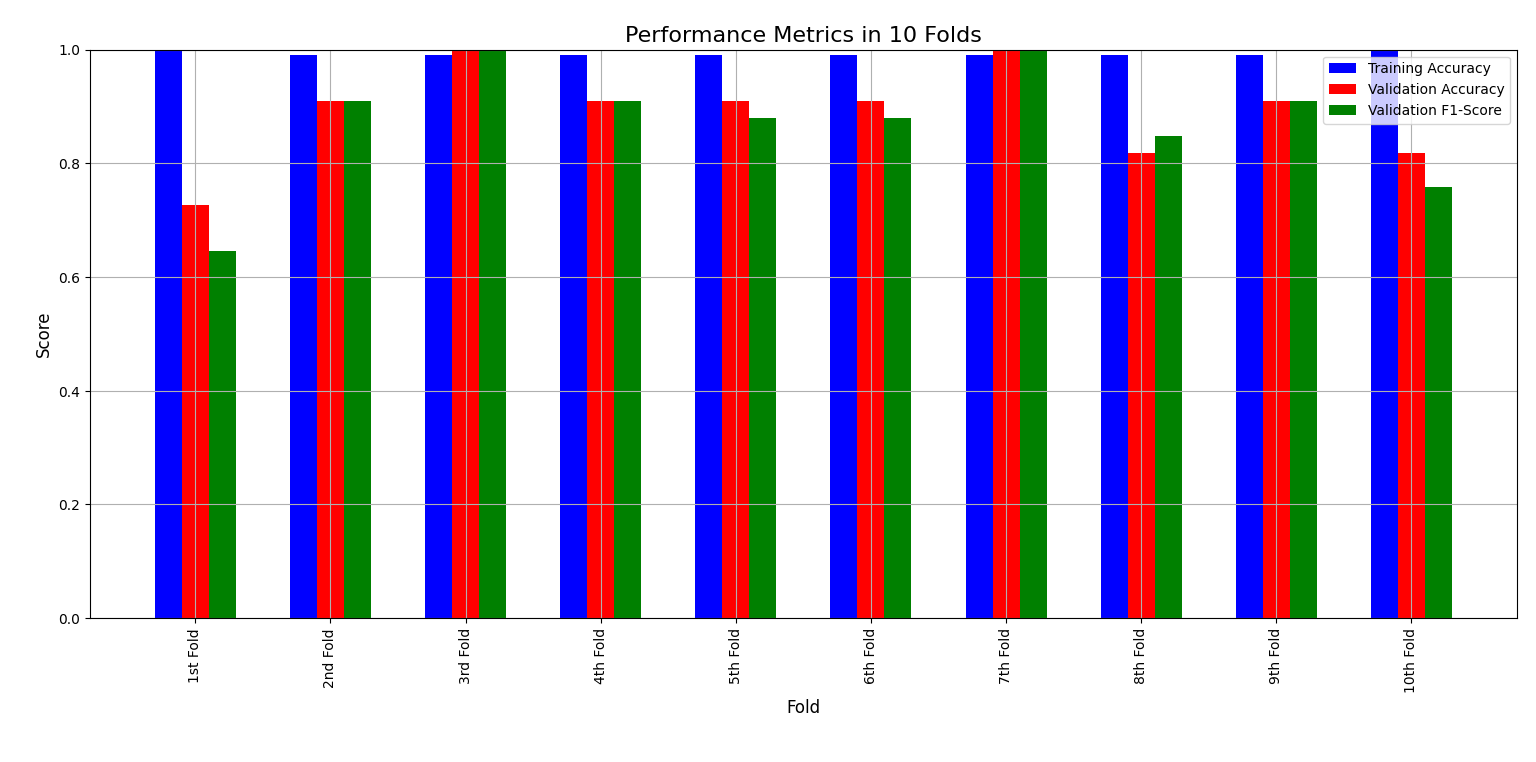}
  \caption{Performance metrics for alphabet signs.}
  \label{fig:3}
\end{figure*}

\begin{figure*}
  \centering
  \includegraphics[width=0.7\textwidth]{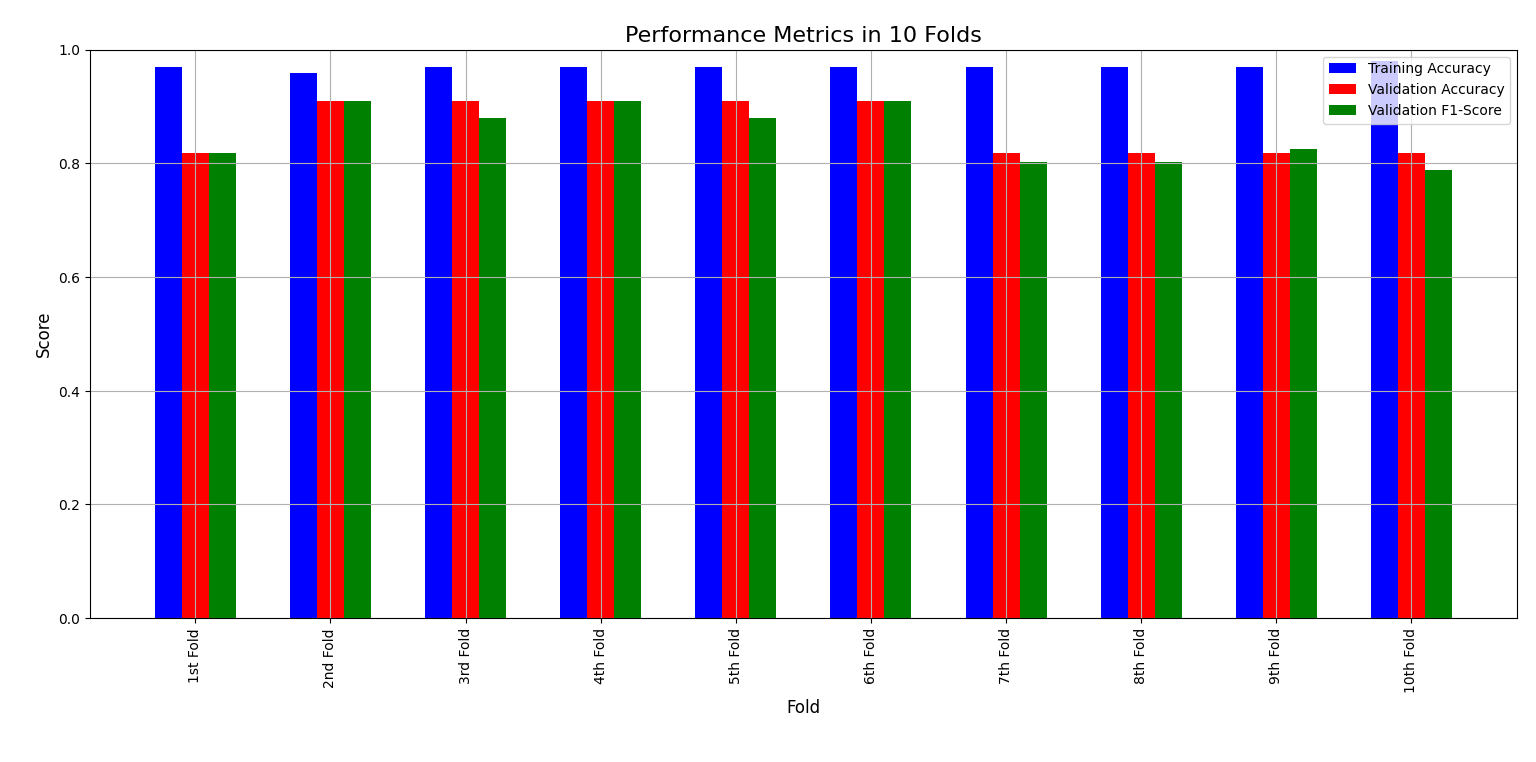}
  \caption{Performance metrics for all sign variations: alphabet letters, words and sentences.}
  \label{fig:4}
\end{figure*}

Figs.~\ref{fig:3} and~\ref{fig:4} respectively show the classification results for the alphabet signs and for all sign variations stored in the dataset. The performance metrics considered are Accuracy as a measure of the overall model precision  in correctly classifying instances and F1-Score as a single measure of the model's performance by balancing both false positives and false negatives. Results are summarized in Table~\ref{tab:1}.

\begin{table}[htbp]
\centering
\caption{Classification Results}
\label{tab:classification_results}
\begin{tabular}{|c|c|c|c|}
\hline
\textbf{Dataset} & \textbf{Metric} & \textbf{Mean Train} & \textbf{Mean Test} \\
\hline
\multirow{2}{*}{Alphabet} & Accuracy & 0.991 & 0.890 \\
\cline{2-4}
& F1-Score & 0.991 & 0.873 \\
\hline
\multirow{2}{*}{All Signs} & Accuracy & 0.969 & 0.863 \\
\cline{2-4}
& F1-Score & 0.967 & 0.852 \\
\hline
\end{tabular}
\label{tab:1}
\end{table}

The classification results reveal that the classification of alphabet signs exhibits higher Accuracy and F1-Score compared to the classification of the other sign variations in the dataset. This indicates that the model achieves greater precision and performs better overall when classifying alphabet signs. However, it is worth noting that both classifications demonstrate robust performance across the training and validation sets, surpassing an 85\% threshold for both Accuracy and F1-Score. These results indicate that the model is able to generalize well to unseen data and maintain consistent performance across different variations of signs.





\section{Conclusion and Future Works} 
In conclusion, the developed sign language recognition algorithm has shown promising performance in accurately recognizing static and dynamic simple signs (from letters to sentences), by utilizing the distances between joints of the skeleton's bones as features. This achievement highlights the potential of incorporating VR technology and AI in sign language education, providing an immersive and interactive learning experience.

However, it is important to acknowledge certain limitations and challenges encountered during the implementation of the algorithm. One such difficulty is the reliance on the Oculus device for accurate hand tracking. The algorithm requires a clear and unobstructed view of the user's hand, imposing restrictions on the testing environment (in terms of light and obstacles). Further optimizations and advancements are needed to ensure robust performance in various real-world scenarios.

Moving forward, several areas can be explored to enhance the functionality and applicability of the sign language recognition application. One potential avenue is the implementation of a mirror effect, allowing the application to recognize the hands of another person facing the user or the user's own hands while touching his/her face. This extension would facilitate interactive communication and learning between sign language users and non-users.

Another important aspect for future development is the continual expansion of the sign language database within the application. 


\section{Acknowledgements}
These results are framed in ISENSE: Innovative Supporting sErvices for uNiversity Students with dEafness project funded by the European Union within the Erasmus+ programme (call 2022), Key Action 2: KA220 Cooperation Partnerships for higher education.
AGREEMENT n. 2022-1-IT02-KA220-HED-000089554. This article has been funded with support from the European Commission.


\begin{thebibliography}{00}

\bibitem{1}
Cheok, M. J., Omar, Z., \& Jaward, M. H. (2019). A review of hand gesture and sign language recognition techniques. International Journal of Machine Learning and Cybernetics, 10(1), 131–153. https://doi.org/10.1007/s13042-017-0705-5
\bibitem{2}
Naglot, D., \& Kulkarni, M. (2016). Real time sign language recognition using the leap motion controller. https://doi.org/10.1109/inventive.2016.7830097
\bibitem{3}
De Luca, V., Gatto, C., Liaci, S., Corchia, L., Chiarello, S., Faggiano, F., Sumerano, G., \& De Paolis, L. T. (2023). Virtual Reality and Spatial Augmented Reality for Social Inclusion: The “Includiamoci” Project. Information, 14(1), 38. https://doi.org/10.3390/info14010038
\bibitem{VRForGood}
Meta, VR For Good, 2023, https://about.meta.com/community/vr-for-good/
\bibitem{4}
Wasserman, B. A., Prate, D., Purnell, B., Muse, A., Abdo, K., Day, K., \& Boyd, L. (2019). vrSensory: Designing Inclusive Virtual Games with Neurodiverse Children. https://doi.org/10.1145/3341215.3356277
\bibitem{5}
Teófilo, M., Lourenço, A., Postal, J., \& De Lucena, V. F. (2018). Exploring Virtual Reality to Enable Deaf or Hard of Hearing Accessibility in Live Theaters: A Case Study. In Lecture Notes in Computer Science (pp. 132–148). Springer Science+Business Media. https://doi.org/10.1007/978-3-319-92052-8-11
\bibitem{6} 
Pigou, L., Dieleman, S., Kindermans, P., \& Schrauwen, B. (2014). Sign Language Recognition Using Convolutional Neural Networks. Springer EBooks, 572–578. https://doi.org/10.1007/978-3-319-16178-5-40
\bibitem{7}
Lang, S. C., Block, M., Rojas, R. (2012). Sign Language Recognition Using Kinect. In Lecture Notes in Computer Science (pp. 394–402). Springer Science+Business Media. https://doi.org/10.1007/978-3-642-29347-4-46
\bibitem{8}
Ghoul, O. E., \& Othman, A. (2022). Virtual reality for educating Sign Language using signing avatar: The future of creative learning for deaf students. 2022 IEEE Global Engineering Education Conference (EDUCON). https://doi.org/10.1109/educon52537.2022.9766692
\bibitem{9} 
Vaitkevičius, A., Taroza, M., Blažauskas, T., Damaševičius, R., Maskeliunas, R., \& Woźniak, M. (2019). Recognition of American Sign Language Gestures in a Virtual Reality Using Leap Motion. Applied Sciences, 9(3), 445. https://doi.org/10.3390/app9030445
\bibitem{10}
Meta. (2023). Quest 2. Retrieved March 13, 2023, from \url{https://www.meta.com/it/quest/products/quest-2/}

\bibitem{11}
Bhushan, S., Alshehri, M., Keshta, I., Chakraverti, A. K., Rajpurohit, J., \& Abugabah, A. (2022). An Experimental Analysis of Various Machine Learning Algorithms for Hand Gesture Recognition. Electronics, 11(6), 968. https://doi.org/10.3390/electronics11060968
\bibitem{12}
Devaraj, A., \& Nair, A. K. (2020). Hand Gesture Signal Classification using Machine Learning. International Conference on Communication and Signal Processing. https://doi.org/10.1109/iccsp48568.2020.9182045
\bibitem{13}
Alnaim, N., Abbod, M. F., \& Albar, A. (2019). Hand Gesture Recognition Using Convolutional Neural Network for People Who Have Experienced A Stroke. 2019 3rd International Symposium on Multidisciplinary Studies and Innovative Technologies (ISMSIT). 
https://doi.org/10.1109/ismsit.2019.8932739
\bibitem{14}
Safavian, S., \& Landgrebe, D. A. (1991). A survey of decision tree classifier methodology. IEEE Transactions on Systems, Man, and Cybernetics, 21(3), 660–674. https://doi.org/10.1109/21.97458
\bibitem{15}
Spread the Sign. Thomas Lydell. European Sign Language Center. Retrieved June 2, 2023, from \url{https://www.spreadthesign.com/}

\end{thebibliography}
\end{document}